\documentclass{moriond}
\pdfoutput=1 

\def\be{\begin{equation}}
\def\ee{\end{equation}}
\def\bea{\begin{eqnarray}}
\def\eea{\end{eqnarray}}

\newcommand{\Photo}{}

\begin{document}
\vspace*{4cm}
\title{RECENT NA48/2 AND NA62 RESULTS}

\author{
D.~MADIGOZHIN%
         \footnote{for the NA48/2 and NA62 Collaborations:
F.~Ambrosino, A.~Antonelli, G.~Anzivino, R.~Arcidiacono,
W.~Baldini, S.~Balev, J.R.~Batley, M.~Behler, S.~Bifani
% ...
}
}

\address{
Joint Institute for Nuclear Research, Dubna, Russia \\
E-mail: madigo@mail.cern.ch
}

\maketitle\abstracts{
The NA48/2 Collaboration at CERN has accumulated and analysed unprecedented
statistics of rare kaon decays in the $K_{e4}$ modes: $K_{e4}(+-)$ ($K^\pm \to \pi^+ \pi^- e^\pm \nu$)
and $K_{e4}(00)$ ($K^\pm \to \pi^0 \pi^0 e^\pm \nu$) with nearly one percent 
background contamination. It leads to the improved measurement of branching 
fractions and detailed form factor studies. New final results from the analysis of 381 
$K^\pm \to \pi^\pm \gamma \gamma$ rare decay candidates collected by 
the NA48/2 and NA62 experiments at CERN are presented. The results include a 
decay rate measurement and fits to Chiral Perturbation Theory (ChPT) description.
}

\section{Introduction}
\label{intro}

The main goal of NA48/2 experiment was the search for CP-violating asymmetry in $K^\pm \to 3\pi^\pm$ 
decays \cite{Batley:2007}. Also it has provided in 2003-2004 a large data sample for charged kaon rare decay studies.
In 2007-2008, the NA62 experiment \cite{Anelli:2005ju} ($R_K$ phase) has collected a large data sample with the
same detector \cite{Fanti:2007vi} but modified beam line.
Two simultaneous $K^+$ and $K^-$ beams were produced by $400$~GeV/$c$ ~protons 
on a beryllium  target. Particles of opposite charge with a central momentum of 
$60$~GeV/$c$ and a momentum band of $\pm 3.8\%$ ($rms$) ($74$~GeV/$c$ $\pm 1.9\%$ for NA62 $R_K$ phase) 
were selected by the system of magnets and collimators. Charged products of $K^\pm$ decays were measured by the 
magnetic spectrometer consisting of four drift chambers (DCH1--DCH4) and a dipole magnet located between DCH2 and DCH3. 
The spectrometer was followed by a scintillator hodoscope. A Liquid Krypton calorimeter (LKr) was used to measure 
the energy of electrons and photons.

\section{$K^{+-}_{e4}$ decay}
\label{kpme4}

Kinematics of the $K^{\pm} \rightarrow \pi^+ \pi^- e^{\pm} \nu$ ($K^{+-}_{e4}$) 
decay is defined by five variables \cite{Cabibbo:1965zz}:
the squared invariant masses of dipion ($S_{\pi}$) and dilepton ($S_e$), the angle 
$\theta_{\pi}$ of $\pi^{\pm}$ in dipion rest frame with respect to the flight direction 
of dipion in the kaon center of mass system,  the similar angle $\theta_e$ of $e^{\pm}$ in 
dilepton rest frame, and the angle $\phi$ between dipion and dilepton planes.

$K_{e4}$ decay amplitude is a product of the leptonic weak current and (V-A)
hadronic current, described in terms of three (F,G,R) axial-vector and one (H) 
vector complex form factors. Sensitivity of $K_{e4}$ decay matrix element to R form 
factor is negligible due to the small mass of electron.
Form-factors may be developed in a partial wave expansion: 
\begin{eqnarray}
\nonumber
F = F_s e^{i \delta_{fs}} +  F_p e^{i \delta_{fp}}  cos \theta_{\pi} + F_d e^{i \delta_{fd}}  cos^2 \theta_{\pi} + ...\\
\nonumber
G = G_p e^{i \delta_{gp}} +  G_d e^{i \delta_{gd}}  cos \theta_{\pi} + ...\\
H = H_p e^{i \delta_{hp}} +  H_d e^{i \delta_{hd}}  cos \theta_{\pi} + ...
\end{eqnarray}

Limiting the expansion to S- and P-waves and considering a unique phase $\delta_p$ for all P-wave
form factors in absence of CP violating weak phases, one will obtain the decay probability, that 
depends only on the real form factor magnitudes $F_s,F_p,G_p,H_p$, a single phase shift 
$\delta=\delta_s-\delta_p$ and kinematic variables.
The form factors can be developed in a series expansion of the dimensionless invariants 
$q^2 = (S_\pi /4m^2_{\pi})-1$ and $S_e/4m^2_{\pi}$ \cite{Amoros:1999mg}.
Two slope and one curvature terms are sufficient to describe the $F_s$ form factor variation 
within the available statistics ($F_s/f_s = 1+(f'_s/f_s) q^2 + (f''_s/f_s) q^4 + (f'_e/f_s) S_e/4m^2_{\pi}$),
while two terms are enough to describe the $G_p$ form factor ($G_p/f_s = g_p/f_s + (g'_p/f_s) q^2$),
and two constants -- to describe the $F_p$ and $H_p$ form factors. 

Hadronic form factors for the S- and P-waves have been obtained 
by NA48/2 concurrently with the phase difference between the S- and P-wave states of $\pi \pi$ 
system, leading to the precise determination of $a^0_0$ and $a^0_2$, the I=0 and I=2 S-wave 
$\pi \pi$ scattering lengths \cite{Batley:2009zz}. 

A high precision measurement of $K^{+-}_{e4}$ form factors and branching fraction
has been published by NA48/2  few years later \cite{Batley:2010zza,Batley:2012rf}. 
$K_{e4}^{+-}$ decay rate was measured relative to $K^{\pm} \to \pi^+\pi^-\pi^{\pm}$ ($K_{3\pi}^{+-}$) 
normalization channel.  A track of charged particle with a momentum $p>2.75$~GeV and $0.9 < E/p < 1.1$ 
was identified as $e^{\pm}$, while the track with $p>5$~GeV and $E/p < 0.8$ was regarded as $\pi^{\pm}$.
A dedicated linear discriminant variable based on shower properties has 
been applied to reject events with one misidentified pion. To suppress $K_{3\pi}^{+-}$ background, the 
vertex invariant mass $M_{3\pi}$ in the $\pi^+\pi^-\pi^{\pm}$ hypothesis and its transverse momentum 
$p_t$ were required to be outside an ellipse centered at PDG kaon mass \cite{Beringer:1900zz} 
and zero transversal momentum, with semi-axes of $20\ MeV/c^2$ and $35\ MeV/c$, respectively. 

The squared missing mass was required to be $ > 0.04\ (Gev/c^2)^2$ to reject 
$\pi^{\pm}\pi^0$ decays with a subsequent $\pi^0 \to e^+e^-\gamma$ process. 
The invariant mass of $e^+e^-$ system was required to be $> 0.03\ GeV/c^2$ in order 
to reject photon conversions. 

For the normalization channel $K_{3\pi}^{+-}$ the $M_{3\pi}$ and $p_t$ were inside a smaller 
ellipse with semi-axes $12\ MeV/c^2$ and $25\ MeV/c$, respectively.
A sample of about 1.11 million $K_{e4}^{+-}$ candidates and about 19 millions of prescaled 
$K_{3\pi}$ candidates were selected from data recorded in 2003-2004.

Two main background sources are known: $K^{\pm} \rightarrow \pi^+ \pi^- \pi^\pm$
decays with subsequent $\pi \rightarrow e\nu$ decay or a pion mis-identified
as an electron; and $K^{\pm} \rightarrow \pi^0 (\pi^0) \pi^\pm$ with subsequent 
$\pi^0 \rightarrow e^+e^-\gamma$ decay with undetected photons and an 
electron mis-identified as a pion. Their admixture in the signal events is estimated to be below 1\%.  

A detailed GEANT3-based \cite{Brun:1994aa} Monte Carlo simulation was used to take into account full 
detector geometry, DCH alignment, local inefficiencies and beam properties. 

The resulting $K_{e4}^{+-}$ branching fraction \cite{Batley:2012rf}
$BR(K_{e4}^{+-}) = (4.257 \pm 0.004_{stat} \pm 0.016_{syst} \pm 0.031_{ext}) 10^{-5}$ is 3 times
more precise than available PDG value \cite{Beringer:1900zz} . It has been used to extract 
the common normalization form factor $f_s$ \cite{Batley:2012rf}.

\section{$K^{00}_{e4}$ decay}
\label{k00e4}

The $K_{e4}^{00}$ rate is measured relative to the $K^{\pm} \to \pi^0\pi^0\pi^{\pm}$ ($K_{3\pi}^{00}$) 
normalization channel. These two modes are collected using the same trigger and with a similar 
event selections. The separation between them occurs only at a later stage.

Events with at least four $\gamma$, detected by LKr, and at least one track, reconstructed from 
spectrometer data, were regarded as $K_{e4}^{00}$ or $K_{3\pi}^{00}$ candidates.
Every combination of 4 reconstructed $\gamma$ with energies $E > 3$~GeV was considered as 
a possible pair of $\pi^0$ decays. 
Reconstructed longitudinal positions $Z_1$ and $Z_2$ of both $\pi^0 \to 2\gamma$ decay 
candidates were required to coincide within $500\ cm$, with their average 
position $Z_n = (Z_1+Z_2)/2$ inside the fiducial volume 106 m long. 

Decay longitudinal position $Z_{ch}$, assigned to the track, 
was defined by the closest distance approach between the track and the beam axis. 
Combined vertex, composed 
of four LKr clusters and one charged track with momentum $p > 5 GeV$, was required to have the 
difference $|Z_n - Z_{ch}|$ less than $800\ cm$. 
If several combinations satisfy the vertex criteria, the case of minimum 
$(\frac{Z_1 - Z_2}{\sigma_n})^2+(\frac{Z_n - Z_{ch}}{\sigma_c})^2$ has been chosen, 
where $\sigma_n$ and $\sigma_c$ are the $Z_n$-dependent widths of corresponding 
distributions.

A track was preliminarily identified as $e^\pm$, if it has an associated LKr cluster with $E/p$ between 0.9 and 1.1,
otherwise $\pi^\pm$ was assumed at the first stage. Further suppression of  pions mis-identified as electrons
is obtained by means of discriminant variable which is a linear combination of $E/p$, shower width and 
energy weighted track-to-cluster distance at LKr front face.

$K_{e4}^{00}$ and $K_{3\pi}^{00}$ decays were discriminated by means of elliptic cuts in the 
($M_{\pi^0\pi^0\pi^{\pm}}, p_t$) plane, where $M_{\pi^0\pi^0\pi^{\pm}}$ is the 
invariant mass of combined vertex in the $K_{3\pi}^0$ hypothesis, and $p_t$ is the transversal momentum.
Elliptic cut separates about 94 million $K_{3\pi}^{00}$ normalization events from about
65000 $K_{e4}^{00}$ candidates. Residual fake-electron background is about 0.65\% of $K_{e4}^{00}$
amount. Background from $K_{3\pi}^{00}$ with the subsequent $\pi^{\pm} \to e^{\pm} \nu$ is 
0.12\% of the signal, and the accidental-related background is about 0.23\%. It gives in total 1\% 
of background admixture.

For the case of $K^{\pm} \rightarrow \pi^0 \pi^0 e^{\pm} \nu$ ($K^{00}_{e4}$) decay, 
due to restrictions of symmetry, matrix element doesn't depend on $\theta_{\pi}$ and $\phi$ angles.
It is parametrized in terms of the only formfactor $F_s$, that
may depend on $S_{\pi}$ and $S_e$. Form factor $F_s$ was extracted from the fit of events 
distribution on ( $S_e , S_{\pi}$) plane, taking into account the acceptance, 
calculated from MC simulation. The following empirical 
parameterization has been used: $F_s/f_s = 1+(f'_s/f_s) q^2 + (f''_s/f_s) q^4 + (f'_e/f_s) S_e/4m^2_{\pi}$ 
for $q^2 > 0$ and $F_s/f_s = 1+ d \sqrt{|q^2/(1+q^2)|}   + (f'_e/f_s) S_e/4m^2_{\pi}$ for $q^2 < 0$.

The results are in agreement with NA48/2 $K_{e4}^{+-}$ analysis described above: 
$f'_s/f_s = 0.149 \pm 0.033_{stat} \pm 0.014_{syst}$, 
$f''_s/f_s = -0.070 \pm 0.039_{stat} \pm 0.013_{syst}$, 
$f'_e/f_s = 0.113 \pm 0.022_{stat} \pm 0.007_{syst}$, 
$d = -0.256 \pm 0.049 \pm 0.016_{syst}$.

The obtained form factor was used to obtain the final result of branching fraction measurement: 
$Br(K_{e4}^{00})=(2.552 \pm 0.010_{stat} \pm 0.010_{syst} \pm 0.032_{ext})10^{-5}$.
It is 10 times more precise, than PDG corresponding value \cite{Beringer:1900zz}. 
Systematic error includes the contributions from background, simulation statistical error, sensitivity to 
form factor, radiation correction, trigger efficiency and beam geometry.
External error comes from uncertainty of normalization channel $K_{3\pi}^{00}$ branching fraction.

\begin{figure}
\begin{center}
\setlength{\unitlength}{1mm}
\resizebox{0.70\columnwidth}{!}{%
\begin{picture}(100.,40.)                
\includegraphics[width=100mm]{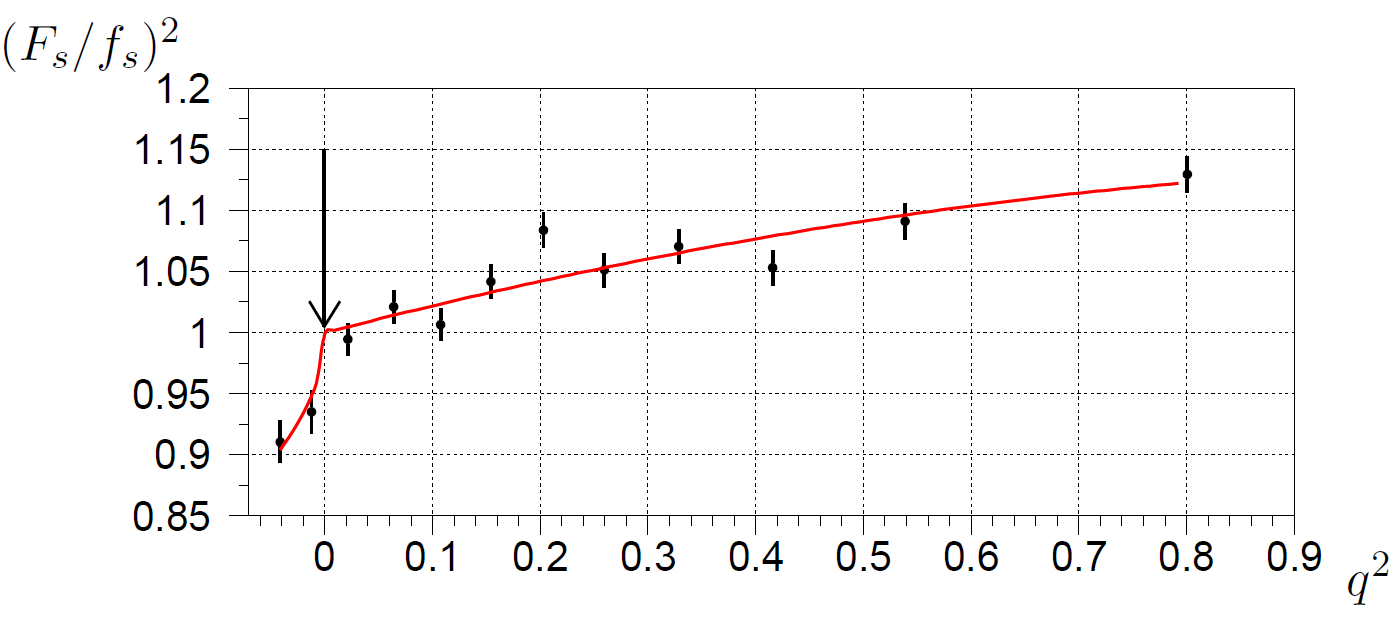}
\end{picture}
}
\end{center}
\caption{ $K_{e4}^{00}$ normalized form factor squared as a function of $q^2$. 
The line corresponds to the adopted empirical fit. The arrow points to the $2 m_{\pi}$ threshold.}
\label{fss}
\end{figure}

Below the threshold of $S_\pi = (2m_{\pi^\pm})^2$ the measured $K_{e4}^{00}$ decay form factor shows
a deficit of events, that is well described by the present empirical parameterization (Fig. \ref{fss}). It is similar 
to the effect of $\pi^+\pi^- \to \pi^0\pi^0$ rescattering in $K^\pm \to \pi^0\pi^0\pi^\pm$ decay 
(cusp effect \cite{Cabibbo:2004gq}), investigated by NA48/2 collaboration earlier \cite{Batley:2009zz} 
on the basis of ChPT formulations.

\section{$K^\pm\to\pi^\pm\gamma\gamma$ decay}

In the ChPT framework, the $K^\pm\to\pi^\pm\gamma\gamma$ decay receives two 
non-interfering contributions at lowest non-trivial order ${\cal O}(p^4)$: the pion and kaon {\it loop amplitude} depending on 
an unknown ${\cal O}(1)$ constant $\hat{c}$ representing the total contribution of the counterterms, 
and the {\it pole amplitude}~\cite{ec88}. 

New measurements of this decay have been performed using data collected during a 3-day special NA48/2 
run in 2004 and a 3-month NA62 run in 2007.
Signal events are selected in the region of $z=(m_{\gamma\gamma}/m_K)^2>0.2$ to reject the $K^\pm\to\pi^\pm\pi^0$ 
background peaking at $z=0.075$.  149 (232) decays candidates are observed in the 2004 (2007) data set, with backgrounds 
contaminations of 10.4\% (7.5\%) from $K^\pm\to\pi^\pm\pi^0(\pi^0)(\gamma)$ decays with merged photon clusters in the 
electromagnetic calorimeter.

The values of $\hat{c}$ in the frameworks of the ChPT ${\cal O}(p^4)$ and ${\cal O}(p^6)$ 
parameterizations \cite{da96} as well as branching ratio have been measured using likelihood fits to the data. 
The main systematic effect is due to the background uncertainty. Uncertainties 
related to trigger, particle identification, acceptance and accidental effects found to be negligible.
The final combined results based on 2004 and 2007 runs data \cite{pigg1,pigg2} are:
$\hat{c}$ for ${\cal O}(p^4)$ fit = $1.72\pm 0.20_{stat} \pm 0.06_{syst}$;
$\hat{c}$ for ${\cal O}(p^6)$ fit =  $1.86 \pm 0.23_{stat} \pm 0.11_{syst}$; 
branching fraction $Br(K_{\pi\gamma\gamma})$ for ${\cal O}(p^6)$ fit = $(1.003\pm0.056)\times 10^{-6}$. 
The model-independent branching ratio for $z > 0.2$ is equal to $(0.965\pm0.063)\times 10^{-6}$.
New results are in agreement with the earlier (based on 31 events) BNL E787 \cite{bnle787} ones.


\section*{References}

\begin{thebibliography}{99}


\bibitem{Batley:2007}
J.R.Batley, et~al., 
Eur. Phys. J. C 52 (2007) 875--891.

\bibitem{Anelli:2005ju}
G.~Anelli, et~al., 
  CERN-SPSC-2005-013. 

\bibitem{Fanti:2007vi}
V.~Fanti, et~al., 
Nucl.Instrum.Meth. A574 (2007) 433--471.

\bibitem{Cabibbo:1965zz}
N.~Cabibbo, A.~Maksymowicz, 
Phys.Rev. 137 (1965) B438--B443.

\bibitem{Amoros:1999mg}
G.~Amoros, J.~Bijnens, 
J.Phys.G G25 (1999) 1607--1622.

\bibitem{Cabibbo:2004gq}
N.~Cabibbo, 
Phys.Rev.Lett. 93 (2004) 121801.

\bibitem{Batley:2009zz}
J.~Batley, et~al.,
Eur.Phys.J. C64 (2009) 589--608.

\bibitem{Batley:2010zza}
J.~Batley, et~al., 
Eur.Phys.J. C70 (2010) 635--657.

\bibitem{Batley:2012rf}
J.~Batley, et~al., 
Phys.Lett. B715 (2012) 105--115.

\bibitem{Beringer:1900zz}
J.~Beringer, et~al., 
Phys.Rev. D86 (2012) 010001.

\bibitem{Brun:1994aa}
R.~Brun, F.~Carminati, S.~Giani, 
CERN-W-5013.

\bibitem{ec88}
G. Ecker, A. Pich and E. de Rafael, Nucl. Phys. {\bf B303} (1988) 665.

\bibitem{da96}
G. D'Ambrosio and J. Portol\'es, Phys. Lett. {\bf B386} (1996) 403.

\bibitem{pigg1}
J.~Batley, et~al., Phys. Lett. {\bf B730} (2014) 141.

\bibitem{pigg2}
J.~Batley, et~al., Phys. Lett. {\bf B732} (2014) 65.

\bibitem{bnle787}
P. Kitching et al., Phys. Rev. Lett. 79 (1997) 4079.

\end{thebibliography}
\end{document}